\documentclass[ACS,STIX1COL]{WileyNJD-v2}
\articletype{ORIGINAL ARTICLE}

\usepackage{color}
\usepackage{graphicx}
\usepackage{dcolumn}
\usepackage{array}
\usepackage{lipsum}
\usepackage{bm}
\usepackage{subfigure}
\usepackage{multirow}
\usepackage{tabularx}
\usepackage{braket}
\usepackage{csquotes}
\graphicspath{{plots/}}
 \usepackage{lipsum}
	
\newcommand{\beq}{\begin{equation}}
\newcommand{\eeq}{\end{equation}}
\newcommand{\bea}{\begin{eqnarray}}
\newcommand{\eea}{\end{eqnarray}}

\renewcommand{\vec}[1]{\mathbf{#1}}
\DeclareMathOperator{\Tr}{Tr}

\begin{document}
\title{Reweighting scheme for the calculation of grand-canonical expectation values in quantum Monte Carlo simulations with a fermion sign problem}

\author[1,2,3]{Paul Hamann}
\author[2]{Jan Vorberger}
\author[2,3]{Tobias Dornheim}

\address[1]{\orgdiv{Institut f\"ur Physik}, \orgname{Universit\"at Rostock}, \orgaddress{D-18057 Rostock, \country{Germany}}}
\address[2]{\orgname{Helmholtz-Zentrum Dresden-Rossendorf}, \orgaddress{\state{Bautzner Landstra{\ss}e 400,  D-01328 Dresden}, \country{Germany}}}
\address[3]{\orgname{Center for Advanced Systems Understanding (CASUS)}, \orgaddress{\state{G\"orlitz}, \country{Germany}}}

\corres{*\email{p.hamann@hzdr.de}}


\abstract{\emph{Ab initio} path integral Monte Carlo (PIMC) simulations constitute the gold standard for the estimation of a broad range of equilibrium properties of a host of interacting quantum many-body systems spanning conditions from ultracold atoms to warm dense quantum plasmas.
A key practical limitation is given by the notorious fermion sign problem, which manifests as an exponential computational bottleneck with respect to system size and inverse temperature. In practice, the sign problem is particularly severe in the grandcanonical ensemble, where the bosonic and fermionic configuration spaces differ not only with respect to the symmetry of the thermal density matrix but, crucially, also with respect to the particle number distribution for a given chemical potential $\mu$ [T.~Dornheim, \textit{J.~Phys.~A}~\textbf{54}, 335001 (2021)]. Here, we present a simple reweighting scheme that basically allows one to retain access to grandcanonical expectation values at the cost of fermionic PIMC simulations in the canonical ensemble for the largest significant particle number in the fermionic sector. As a practical example, we consider the warm dense electron gas, which has attracted considerable recent attention due to its relevance, e.g., for the modeling of compact astrophysical objects and inertial fusion energy applications.}

\maketitle

\section{Introduction}

The accurate description of interacting quantum many-body systems continues to be the main challenge in a variety of research fields, including many aspects of physics, quantum chemistry and material science.
In thermal equilibrium, \emph{ab initio} quantum Monte Carlo (QMC) methods~\cite{anderson2007quantum,Foulkes_RMP_2001,cep} have emerged as a highly successful concept, which, in principle, allows for an exact description of a given system without the need for any empirical external input such as the exchange--correlation functional in density functional theory~\cite{Goerigk_PCCP_2017}. 
At finite temperatures, the method of choice is often given by the \emph{ab initio} path integral Monte Carlo (PIMC) method~\cite{Berne_JCP_1982,Takahashi_Imada_PIMC_1984,Pollock_PRB_1984,boninsegni1}, which is based on Feynman's imaginary-time path-integral representation of statistical mechanics~\cite{kleinert2009path}. Specifically, PIMC is based on the celebrated classical isomorphism~\cite{Chandler_JCP_1981}, where the complicated quantum many-body system of interest is effectively mapped onto an ensemble of de-facto classical interacting ring-polymers.
The situation becomes somewhat more complex for indistinguishable particles obeying Fermi-Dirac or Bose-Einstein statistics. Here, different ring-polymers are allowed to connect with each other to form so-called \emph{permutation cycles}~\cite{Dornheim_permutation_cycles}, which is effectively being accounted for by modern sampling techniques such as the worm algorithm by Boninsegni \emph{et al.}~\cite{boninsegni1,boninsegni2}.
Indeed, state-of-the-art implementations allow for quasi-exact simulations of $N\sim10^3-10^4$ quantum particles~\cite{boninsegni1,Dornheim_JPCL_2024,svensson2025acceleratedfreeenergyestimation}, and corresponding PIMC-based investigations have been pivotal for our understanding of a variety of important physical effects such as superfluidity~\cite{cep,Sindzingre_PRL_1989,Dornheim_PRA_2020,Yan_Blume_PRL_2014}, collective and pair excitations~\cite{Saccani_Supersolid_PRL_2012,Filinov_PRA_2012,Filinov_PRA_2016,dornheim_dynamic,Chuna_JCP_2025,Ferre_PRB_2016}, as well as crystallization~\cite{Clark_PRL_2009,Filinov_PRL_2001,Bhattacharya2016,Boninsegni_PRA_2013}.

An important limitation of the PIMC method is the notorious fermion sign problem~\cite{troyer,Loh_PRB_1990,dornheim_sign_problem}, which leads to an exponential increase in the required compute time with respect to system parameters such as the system size $N$ and the inverse temperature $\beta=1/k_\textnormal{B}T$ in PIMC simulations of quantum degenerate fermions. While the pressing need to understand interacting Fermi-systems has sparked a remarkable surge of new developments over the last decade or so~\cite{Brown_PRL_2013,Schoof_PRL_2015,Schoof_CPP_2015,Blunt_PRB_2014,Malone_JCP_2015,Malone_PRL_2016,Joonho_JCP_2021,Yilmaz_JCP_2020,Dornheim_NJP_2015,Hirshberg_JCP_2020,Dornheim_JCP_2020,Dornheim_PRB_2025,Chin_PRE_2015,Rubenstein_JCTC_2020,dornheim_prl,Xiong_JCP_2022,morresi2024normalliquid3hestudied,Xiong_PRE_2023,Dornheim_NatComm_2025}, the sign problem still substantially limits the range of applicability of PIMC to moderate levels of quantum degeneracy.
This situation is particularly severe in the grandcanonical ensemble (GCE), where the number of particles in the simulation cell is allowed to fluctuate.
While its implementation is algorithmically relatively straightforward~\cite{boninsegni1,boninsegni2}, PIMC simulations with a sign problem are being carried out in a bosonic reference system in practice, and the correct fermionic expectation values of interest are being extracted subsequently from the cancellation of positive and negative terms; this is the root cause of the fermion sign problem.
In the GCE, this situation is exacerbated by the different particle number distribution functions of bosons and fermions for the same value of $\mu$.

\begin{figure}\centering
\includegraphics[width=0.7\textwidth]{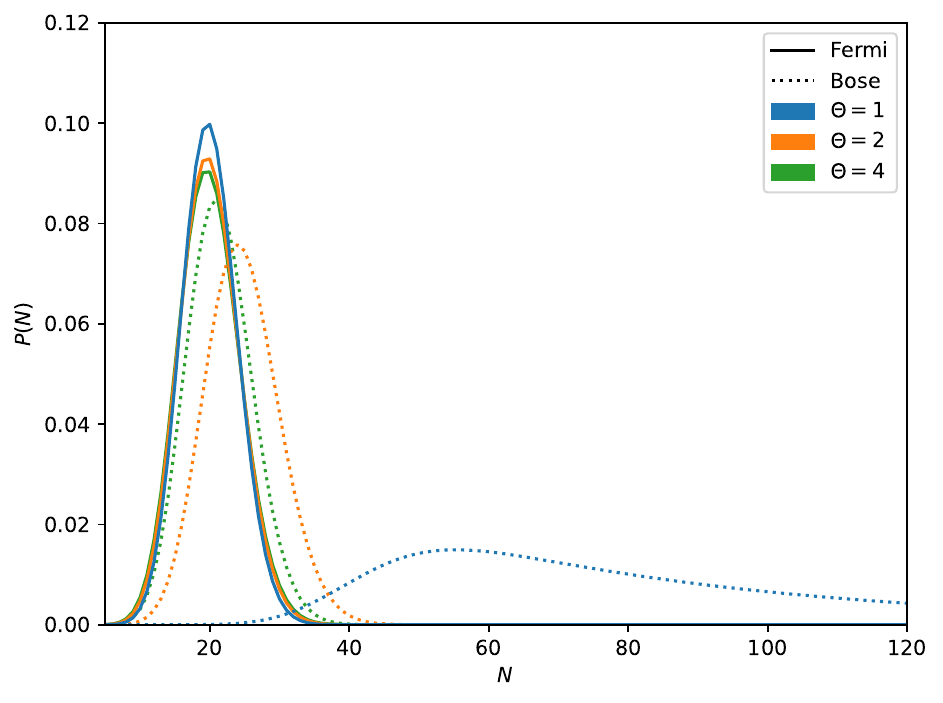}
\caption{\label{fig:Ndist} Particle number distribution in grand-canonical PIMC simulations of $\langle N\rangle = 20$ non-interacting fermions (same spin) at different values of the degeneracy parameter $\Theta$. }
\end{figure}

The associated growing inefficiency of sampling fermionic expectation values based on the bosonic weight function 
is illustrated in Fig.~\ref{fig:Ndist}, which shows the diminishing overlap between fermionic (solid) and bosonic (dotted) particle number distributions of an ideal Fermi gas as the system becomes more degenerate and effects of quantum statistics become important. Here parameters are chosen such that the mean fermionic particle number $\langle N\rangle_F = 20$ is kept constant for different values of the degeneracy parameter $\Theta = k_B T/E_F$. 
In the weakly quantum degenerate regime ($\Theta=4$, green curves), the bosonic and fermionic distribution functions are very similar and the PIMC sampling in the bosonic reference systems is efficient.
Decreasing the temperature by a factor of two ($\Theta=2$, yellow curves) substantially shifts the bosonic curve to larger particle numbers, which becomes problematic in two ways: i) a lot of simulation time is wasted at large $N$, which are not relevant for the fermionic expectation value of interest; all associated contributions will approximately cancel due to the sign problem; ii) configurations with small $N$, which are of substantial importance for the Fermi-Dirac case, are hardly explored in the bosonic reference calculation, leading to an undersampling and, consequently, a large statistical uncertainty.
At $\Theta=1$ (blue curves), which is often still considered as a moderate degree of quantum degeneracy, the bosonic and fermionic particle number distribution are strikingly dissimilar. 
Indeed, it is immediately obvious that most of the simulation time in the bosonic reference system will be wasted as the overlap between the two types of particle statistics basically vanishes. 
The associated difficulty of fermionic PIMC simulations in the GCE thus strongly hampers the calculation of grandcanonical observables such as the compressibility and the Matsubara Green function.

In this work, we present an easy workaround to the grandcanonical exacerbation of the fermion sign problem~\cite{Dornheim_JPA_2021} in PIMC. Specifically, we propose a re-weighting procedure that allows for the computation of grandcanonical expectation value at the computation cost of PIMC simulations in the canonical ensemble. 
As a practical ensemble, we consider two representative systems: i) the ideal Fermi gas, for which exact reference results are available both for finite~\cite{Zhou_2018} and infinite systems~\cite{quantum_theory} and ii) the uniform electron gas (UEG)~\cite{review,loos}, which is often considered as the archetypical model system for interacting electrons.

The paper is organized as follows: Sec.~\ref{sec:theory} deals with the required theoretical background, starting with a discussion of the UEG model and its parameters (\ref{sec:system}) and a brief introduction to fermionic PIMC simulations (\ref{sec:PIMC}). After discussing the computation of grand-canonical observables, with and without reweighting (\ref{sec:GCE}) this section is concluded by giving the necessary formulas for obtaining exact results for the non-interacting system (\ref{ss:exact}). These are subsequently used as a reference in Sec.~\ref{sec:results}, where the reweighting scheme is first tested on PIMC results for the free electron gas at varying degeneracy (\ref{sec:ideal}), before applying it to the interacting UEG (\ref{sec:interacting}).

\section{Theory}\label{sec:theory}

We assume Hartree atomic units throughout this work.

\subsection{System parameters and Hamiltonian\label{sec:system}}

Both the UEG and the ideal Fermi gas are conveniently characterized by a small set of dimensionless parameters~\cite{Ott2018,vorberger2025roadmapwarmdensematter}: (i) the density parameter $r_s=(3/4\pi n)^{1/3}$, also known as Wigner-Seitz radius in the literature, also serves as the quantum coupling parameter, with $n=N/\Omega$ being the total electronic number density for $N$ electrons in a volume $\Omega$. In the high-density limit of $r_s\to0$, the UEG becomes an ideal Fermi gas, whereas it forms a strongly correlated electron liquid~\cite{dornheim_electron_liquid,Takada_PRB_2016,Chuna_JCP_2025,dornheim_dynamic,Tolias_JCP_2021,Tolias_JCP_2023} and, eventually, even a Wigner crystal~\cite{Drummond_PRB_Wigner_2004,Azadi_Wigner_2022} for large values of $r_s$ when the level of quantum degeneracy is being kept constant; (ii) the degeneracy temperature $\Theta=k_\textnormal{B}T/E_\textnormal{F}$, where $E_\textnormal{F}$ is the Fermi temperature~\cite{quantum_theory}, quantifies the level of quantum delocalization and exchange, with $\Theta\gg1$ and $\Theta\ll1$ corresponding to the semi-classical~\cite{Dornheim_HEDP_2022,Roepke_PRE_2024,tolias2025exactseriesexpansionfrequency} and ground-state limits, respectively; (iii) the spin-polarization parameter $\xi=(N^\uparrow-N^\downarrow)/(N^\uparrow+N^\downarrow)$, where $N^\uparrow$ and $N^\downarrow$ indicate the number of spin-up and spin-down electrons.
For completeness, we note that the degree of spin-polarization $\xi$ must not be confused with the fictitious spin-variable that is usually denoted by the same letter, and which has been applied rather successfully to path integral simulations with a fermion sign problem in the canonical ensemble in a series of recent publications~\cite{Xiong_JCP_2022,Xiong_PRE_2023,Dornheim_JCP_xi_2023,Dornheim_JPCL_2024,Dornheim_JCP_2024,Dornheim_NatComm_2025,Dornheim_MRE_2024,dornheim2025fermionicfreeenergiestextitab,morresi2025studyuniformelectrongas,morresi2024normalliquid3hestudied,Yang_Entropy_2025,svensson2025acceleratedfreeenergyestimation}.

In this work, we focus on the \emph{warm dense matter} regime, which is defined by~\cite{wdm_book,vorberger2025roadmapwarmdensematter,Ott2018} $r_s\sim\Theta\sim1$. These conditions naturally occur in a variety of compact astrophysical objects such as giant planet interiors~\cite{Benuzzi_Mounaix_2014,wdm_book} and brown dwarfs~\cite{becker}. Moreover, warm dense matter is relevant for a number of cutting-edge technological applications such as material science and discovery~\cite{Lazicki2021,Kraus2016,Frost2024} as well as inertial confinement fusion~\cite{hu_ICF,Betti2016,Hurricane_RevModPhys_2023}.
In particular, the absence of any small parameters necessitates the application of state-of-the-art \emph{ab initio} methods such as PIMC to holistically take into account the complex interplay of effects such as Coulomb coupling, quantum degeneracy and delocalization, and strong thermal excitations~\cite{wdm_book,new_POP,Dornheim_review,vorberger2025roadmapwarmdensematter}.

The Hamiltonian of the $N$-particle UEG is given by~\cite{review}
\begin{eqnarray}\label{eq:Hamiltonian}
    \hat{H}_N = \underbrace{-\frac{1}{2}\sum_{l=1}^N\nabla_l^2}_{\hat{K}} + \underbrace{\frac{1}{2}\sum_{l\neq k}^N W_\textnormal{E}(\hat{\mathbf{r}}_l,\hat{\mathbf{r}}_k) + \frac{N}{2}\xi_\textnormal{M}}_{\hat W}\ ,
\end{eqnarray}
with $\hat{K}$ and $\hat{W}$ corresponding to the kinetic and interaction parts, respectively; the ideal Fermi gas corresponds to $\hat{W}\equiv 0$.
We consider a cubic simulation cell of volume $\Omega=L^3$ and implement standard periodic boundary conditions. The electron--electron interaction is given by the usual Ewald sum taking into account the interaction of the infinite array of periodic images, and we follow the convention introduced in Ref.~\cite{Fraser_PRB_1996} where $W_\textnormal{E}(\mathbf{r}_1,\mathbf{r}_2)$ already contains terms due to the uniform neutralizing background. We note that alternative perodic pair potentials such as the spherically averaged version by Yakub and Ronchi~\cite{Yakub2005,Yakub_JCP_2003} have been explored in the literature~\cite{Filinov_PRE_2015,Demyanov_2022,Levashov_CPP_2024,dornheim2025applicationsphericallyaveragedpair,Filinov_PRE_2023}.
Finally, the Madelung constant $\xi_\textnormal{M}$ takes into account self interactions of a charge with its own array of periodic images and its background.

\subsection{Path integral Monte Carlo and the fermion sign problem\label{sec:PIMC}}

In this section, we give a brief overview of the most basic relations that are required in the context of the present work. The interested reader is referred to the ample existing literature on the PIMC method~\cite{Berne_JCP_1982,Pollock_PRB_1984,Takahashi_Imada_PIMC_1984,cep,Marienhagen_JCP_2025} and its implementation~\cite{boninsegni1,Dornheim_PRB_nk_2021} for more exhaustive and pedagogical introductions.

A fundamental property in statistical mechanics is given by the canonical (i.e., particle number $N$, volume $\Omega$, and inverse temperature $\beta=1/k_\textnormal{B}T$ are being kept constant) partition function, which, in coordinate representation, reads
\begin{eqnarray}\label{eq:Z}
    Z(N,\Omega,\beta) = \frac{1}{N!}\sum_{\sigma\in S_N} (-1)^{N_\textnormal{pp}(\sigma)} \int \textnormal{d}\mathbf{R}\ \bra{\mathbf{R}} e^{-\beta\hat{H}}\ket{\hat{\pi}_\sigma\mathbf{R}}\ .\
\end{eqnarray}
Note that we restrict ourselves here to a single species of spin-polarized fermions, with the generalization to multiple particle species such as spin-up and spin-down electrons being straightforward~\cite{Dornheim_PRB_2016}.
The meta-variable $\mathbf{R}=(\mathbf{r}_1,\dots,\mathbf{r}_N)^T$ contains the coordinates of all $N$ particles and the correct anti-symmetry of the fermionic thermal density matrix with respect to the exchange of particle coordinates is ensured by explicitly summing over all possible permutations $\sigma$ of the $N$-body permutation group $S_N$.
Here, the operator $\hat{\pi}_\sigma$ realizes a particular permutation, which, depending on the corresponding number of associated pair exchanges $N_\textnormal{pp}(\sigma)$, can contribute with a positive or negative sign.

In practice, Eq.~(\ref{eq:Z}) generally cannot directly be evaluated as the matrix elements of the density operator $\hat\rho=e^{-\beta\hat{H}}$ are usually unknown.
The combined application of the exact semi-group property of $\hat\rho$ with a Trotterization~\cite{trotter} (which, strictly speaking, only holds for potentials $\hat{W}$ that are bounded from below, see, e.g., Refs.~\cite{kleinert2009path,Bohme_PRE_2023,MILITZER201688}, allows one to avoid this obstacle, leading to the compact symbolic representation of the partition function
\begin{eqnarray}
    Z(N,\Omega,\beta) = \sumint\textnormal{d}\mathbf{X}\ W(\mathbf{X})\ .
\end{eqnarray}
In short, the original quantum many-body system of interest has been mapped onto a set of $P$ particle coordinates $\mathbf{X}=(\mathbf{R}_0,\dots,\mathbf{R}_{P-1})^T$, and the notation $\sumint\textnormal{d}\mathbf{X}$ implies both the integration over all coordinates as well as the sum over all pair permutations $\sigma$.
The important point is that the (unnormalized) weight function $W(\mathbf{X})$ can actually be evaluated in practice.
The basic idea of the PIMC method is then to utilize a suitable modern implementation of the celebrated Metropolis algorithm~\cite{metropolis} to generate a Markov chain of configurations $\mathbf{X}$ that is distributed according to the probability $P(\mathbf{X}) = W(\mathbf{X})/Z(N,\Omega,\beta)$.

For fermions, $P(\mathbf{X})$ can be both positive and negative, preventing its straightforward interpretation as a probability, which is a prerequisite for the Metropolis sampling.
As a practical workaround, one can sample configurations according to the modified probability distribution $P'(\mathbf{X})=|W(\mathbf{X})|/Z'$, with the modified normalization
\begin{eqnarray}\label{eq:Z_prime}
    Z' = \sumint\textnormal{d}\mathbf{X}\ |W(\mathbf{X})|\ ;
\end{eqnarray}
note that we suppress the canonical state parameters $N$, $\Omega$, and $\beta$ in the following as the same reasoning also generalizes to the GCE, which is introduced in Sec.~\ref{sec:GCE} below.
The exact fermionic expectation value of an arbitrary observable $\hat{O}$ of interest is then recovered by 
\begin{eqnarray}\label{eq:ratio}
    \braket{\hat O}_\text{Fermi} = \frac{\braket{\hat O\hat S}_\text{Bose}}{\braket{\hat S}_\text{Bose}}\ ,
\end{eqnarray}
where the sign operator $\hat S$ measures the sign of the true fermionic configuration weight, $S(\mathbf{X})=W(\mathbf{X)}/|W(\mathbf{X})|$.
In the case of the direct PIMC method, which is being considered throughout the present work, Eqs.~(\ref{eq:Z_prime}) and (\ref{eq:ratio}) imply that we compute fermionic observables by sampling a bosonic configuration space and subsequently consider the impact of the fermionic asymmetry by evaluating the signful terms in Eq.~\eqref{eq:ratio}.
The denominator of the latter is usually denoted as the \emph{average sign} $S\equiv \braket{\hat S}_\text{Fermi}$ in the literature~\cite{dornheim_sign_problem,Dornheim_JPA_2021} and constitutes a useful measure for the amount of cancellation of positive and negative terms.
In practice, Monte Carlo error bars scale as $\Delta O/O\sim1/S$, and simulations are generally feasible for $S\gtrsim10^{-2}-10^{-3}$.
For PIMC simulations in the canonical ensemble, it asymptotically holds $S\sim e^{-\beta N \Delta f}$, with $\Delta f$ being the free energy density difference between the bosonic reference system and the actual fermionic system of interest. 
The resulting exponential increase in the compute time that is required to attain a given level of accuracy is the aforementioned \emph{fermion sign problem}, constituting one of the most stifling limitations in a variety of disciplines across physics, quantum chemistry, and material science.

\subsection{Computing grand-canonical expectation values from canonical measurements\label{sec:GCE}}

The worm algorithm samples configurations according to their contribution to the grand canonical partition function, which is given by the sum of canonical partition functions for all possible values of $N$, weighted by the fugacity $e^{\beta\mu N}$:
\begin{eqnarray}\label{eq:Z_gc}
    Z_{GC}(\mu,\Omega,\beta) = \sum_{N=0}^\infty Z_C(N,\Omega,\beta) e^{\beta\mu N}.
\end{eqnarray}
When only looking at configurations with the same particle number $N$, their relative weights still follow from the canonical partition function $Z_C(N,\Omega,\beta)$, which allows us to compute canonical expectation values for all different particle numbers encountered in the same simulation:
\begin{equation}
    \langle \hat{O} \rangle_{N'} = \Tr \hat{\rho}_{N'} \hat{O} = \langle \hat{O} \cdot \delta_{N,N'} \rangle_{GC}.
\end{equation}
Furthermore, we can determine the relative contribution of each $N$-particle sector to the total partition function (particle number distribution) by creating a histogram of the number of particles present in the simulation whenever a measurement is taken:
\begin{equation}\label{eq:pn}
    P(N) = \frac{Z_C(N,\Omega,\beta)}{Z_{GC}(\mu,\Omega,\beta)} e^{\beta\mu N} = \langle \delta_{N,N'} \rangle_{GC}.
\end{equation}
Grand-canonical expectation values can then simply be expressed as:
\begin{equation}\label{eq:A_GC}
    \langle \hat{A} \rangle_{GC} = \sum_N P(N) \langle \hat{O}\rangle_N\ .
\end{equation}
Clearly, Eq.~(\ref{eq:A_GC}) directly implies that we can compute any grand-canonical expectation value at the cost of a set of PIMC simulations in the canonical ensemble if the true fermionic particle number distribution were known. 
Since $Z_{GC}$ only acts as a normalization factor, ensuring that $\sum_N P(N) = 1$, the relative (i.e., unnormalized) weights are all we need to compute the particle number distribution at any given chemical potential. In practice, we thus choose to carry out a PIMC simulation at a chemical potential $\mu$ that is chosen in such a way that the bosonic particle number distribution resembles the true fermionic distribution of interest as much as possible; the unbiased fermionic particle number distribution of interest is then given by
\begin{equation}\label{eq:pn_mu}
    P'(N) = \frac{  P(N) e^{\beta (\mu' - \mu) N} } { \sum_{N'} P(N') e^{\beta (\mu' - \mu)N'}}\ ,
\end{equation}
with $e^{\beta (\mu' - \mu) N}$ acting as a reweighting factor.

To have further control on the particle number distribution which is actually sampled in the simulation, the weight function can be modified to include arbitrary particle number dependent terms, which are then removed in the subsequent reweighting procedure~\cite{Herdman_PRB_2014,dornheim2024chemicalpotentialwarmdense}. A simple choice is given by a harmonic potential centered around a target particle number,  which is referred to as the restricted grand-canonical ensemble:
\begin{equation}\label{eq:restricted}
    Z_{RGC}(\mu,\Omega,\beta) = \sum\limits_{N=0}^\infty Z_C (N,\Omega,\beta) e^{\beta\mu N} e^{ - (N-\bar{N})/2\sigma^2}
\end{equation}
Here, the parameter $\sigma$ controls the steepness of the potential, penalizing deviations from the target particle number $\bar{N}$. For $\sigma\to\infty$, the usual grand-canonical ensemble is recovered. Such a restriction is useful to reduce particle number fluctuations when utilizing the worm algorithm to perform canonical measurements at a certain target density. Here we use it to avoid sampling the large tail of the bosonic particle number distribution at strong degeneracy (c.f. Fig.~\ref{fig:Ndist} for $\Theta=1$).

\subsection{Exact results for non-interacting quantum gases of finite size}\label{ss:exact}
As a starting point, the ideal Fermi gas, i.e. the UEG with all interaction turned off, $\hat{W}\equiv0$ in Eq.~(\ref{eq:Hamiltonian}), will serve as a test case. This is a particularly useful scenario to consider, because it a) provides us with a realistic sign problem scaling similar to that of the fully interacting UEG with regards to particle number and degeneracy (increasing the coupling strength $r_s$ while keeping the degeneracy parameter $\Theta$ constant attenuates the formation of exchange cycles~\cite{Dornheim_permutation_cycles} and leads to the sign problem becoming less severe) while b) still allowing for an exact analytical description, which makes it possible to investigate how well the reweighting procedure performs in parameter regions where direct PIMC simulations become unfeasible.

\begin{figure}[h]
    \centering
    \includegraphics[width=0.6\textwidth]{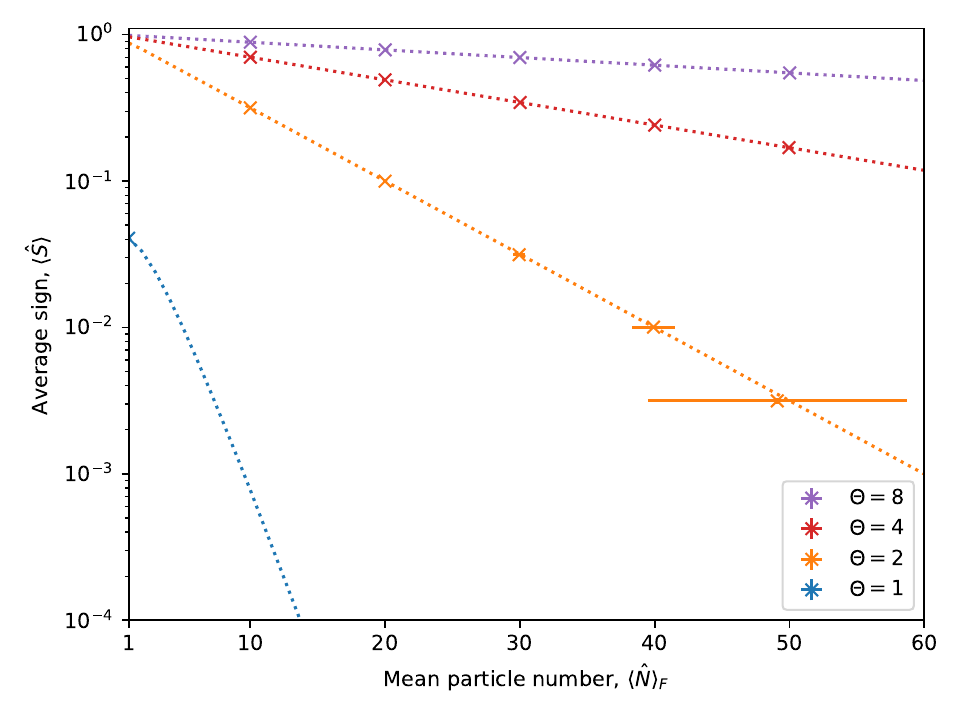}
    \caption{Average sign in grand-canonical PIMC simulations of non-interacting electrons (same spin) in dependence of the system size. Shown are exact analytical predictions (dotted lines) and PIMC results (crosses with error bars) for different choices of the degeneracy parameter $\Theta$.}
    \label{fig:sign}
\end{figure}

Exact results for finite systems are readily available using a decomposition of the partition function into exchange cycles following Refs.~\cite{glaum,Schmidt1998}.
The partition function of a single particle in periodic boundary conditions is given by the Jacobi theta function:
\begin{equation}
    Z_1(\beta) = \sum\limits_{\vec{k}} e^{-\beta \vec{k}^2/2} = \vartheta_3 ( 0; e^{-\frac{\beta}{2} \left(\frac{2\pi}{L}\right)^2})^3,
\end{equation}
where the sum goes over the momentum eigenstates of the simulation cell, i.e. $\vec{k} = 2\pi \vec{n}/L$ with $\vec{n} \in \mathbb{Z}^3$ and $L$ is the length of the box.
Assuming no interaction, the fermionic/bosonic $N$-particle density operator is given by an anti-/symmetrized product. Applying the Laplace expansion to the occurring determinant/permanent leads to the following recursion formula for the $N$-particle partition function 
\begin{equation}\label{eq:Z_exact}
    Z_N^{\text{Bose/Fermi}}(\beta) = \frac{1}{N} \sum\limits_{n=1}^N \left( \pm 1\right)^{n+1} Z_1(n\beta)Z^{\text{Bose/Fermi}}_{N-n}(\beta),
\end{equation}
where the upper/lower sign is for bosons/fermions and $Z_0=1$. 
One-particle observables may be expressed in terms of their value for the single-particle system as follows:
\begin{equation}
    \langle \hat{O}\rangle_N^{\text{Bose/Fermi}} (\beta) = \frac{1}{Z^{\text{Bose/Fermi}}_N(\beta)} \sum\limits_{n=1}^N \left( \pm 1\right)^{n+1} \langle \hat{O}\rangle_1 (n\beta) Z_1(n\beta)Z^{\text{Bose/Fermi}}_{N-n}(\beta)
\end{equation}
Having results for arbitrary $N$ at hand, the grand-canonical partition function for a given chemical potential directly follows from Eq.~\eqref{eq:Z_gc}. The particle number distribution $P(N)$ as already presented in Fig.~\ref{fig:Ndist} is obtained from Eq.~\eqref{eq:pn}. Using results for different $N$ in combination with Eqs.~\eqref{eq:pn} and \eqref{eq:A_GC} allows us to compute expectation values in the grand-canonical ensemble, e.g. for the mean particle number:
\begin{equation}
    \langle N\rangle = \frac{\sum\limits_{N=0}^\infty N e^{\beta\mu N} Z_N} { \sum\limits_{N=0}^\infty e^{\beta\mu N} Z_N}.
\end{equation}
These numerical values for $P(N)$ and $N$ in dependence on the simulation parameters $(\Omega,\beta,\mu)$ will serve as reference throughout the next section. Another quantity of interest is the average sign discussed in Sec.~\ref{sec:PIMC}, which quantifies the rate at which bosonic contributions cancel in the fermionic partition function and can be directly computed from Eq.~\eqref{eq:Z_exact}, i.e. $\langle \hat{S}\rangle_N = Z_N^\text{Fermi}/Z_N^\text{Bose}$, or respectively using the grand-canonical partition function $\langle \hat{S}\rangle_{GC} = Z_{GC}^\text{Fermi} / Z_{GC}^\text{Bose}$.

Results for the average sign encountered in the grand-canonical ensemble are shown in Fig.~\ref{fig:sign} in dependence on the system size (quantified by the average number of electrons in the simulation cell) and for different values of the degeneracy parameter $\Theta$.
In contrast to canonical PIMC simulations, in which the sign by necessity approaches unity if there is only one particle present in the simulation cell as no permutation cycles can be formed (no exchange occurs), configurations of varying particle numbers contribute to the grand-canonical partition function for which $\langle N\rangle =1$ which leads to an average sign smaller than one, which becomes more noticeable at strong degeneracy. 

Similar to the canonical case, the average sign approaches an exponential decay in the large-$N$ limit, with a rate strongly depending on the degeneracy parameter (note the logarithmic axis!). For weakly degenerate systems, permutation cycles are rarely formed. At $\Theta=8$, the average sign still hasn't decayed by one order of magnitude for a mean particle number as large as $\langle N\rangle = 100$ and poses no computational barrier for carrying out simulations. At $\Theta=1$, the average sign reaches a value as low as $10^{-3}$ at a mean particle number of only $\langle N\rangle = 10$, a point at which direct grand-canonical PIMC simulations become practically impossible.

\section{Results}\label{sec:results}

All PIMC results in this work have been obtained using the fully grandcanonical implementation of the worm algorithm by Boninsegni \emph{et al.}~\cite{boninsegni1,boninsegni2} as it has been implemented into the open-source \texttt{ISHTAR} code~\cite{ISHTAR}.
A repository with the PIMC simulation raw data is available online~\cite{repo}.

\subsection{Ideal Fermi Gas\label{sec:ideal}}

\begin{figure}[h]\centering
\includegraphics[width=0.6\textwidth]{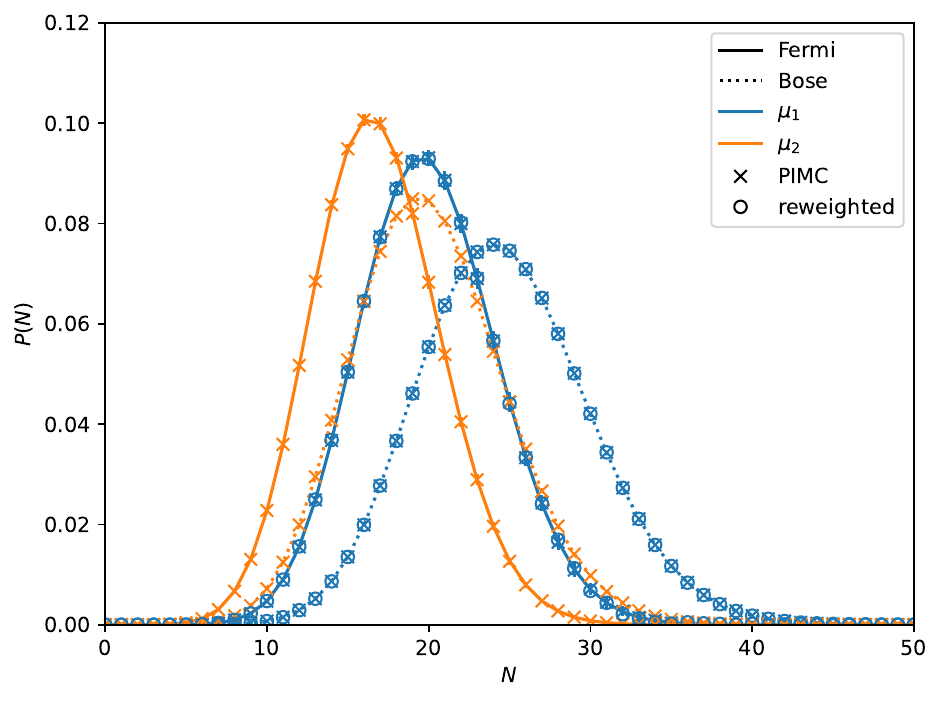}
\caption{Particle number distribution of non-interacting finite quantum gases. Shown are results for two different values of the chemical potential $\mu$ at which $\langle N \rangle = 20$ for bosons/fermions. These are obtained directly from PIMC (crosses) and by reweighting results obtained from the simulation at $\mu_2$ (open circles). For comparison, exact reference values for bosons (fermions) following from Eq.~\eqref{eq:pn} are shown by the dotted (solid) lines.}\label{fig:bose_fermi_ideal_reweighted}
\end{figure} 

\begin{figure}[h]
    \centering
    \includegraphics[width=1.0\textwidth]{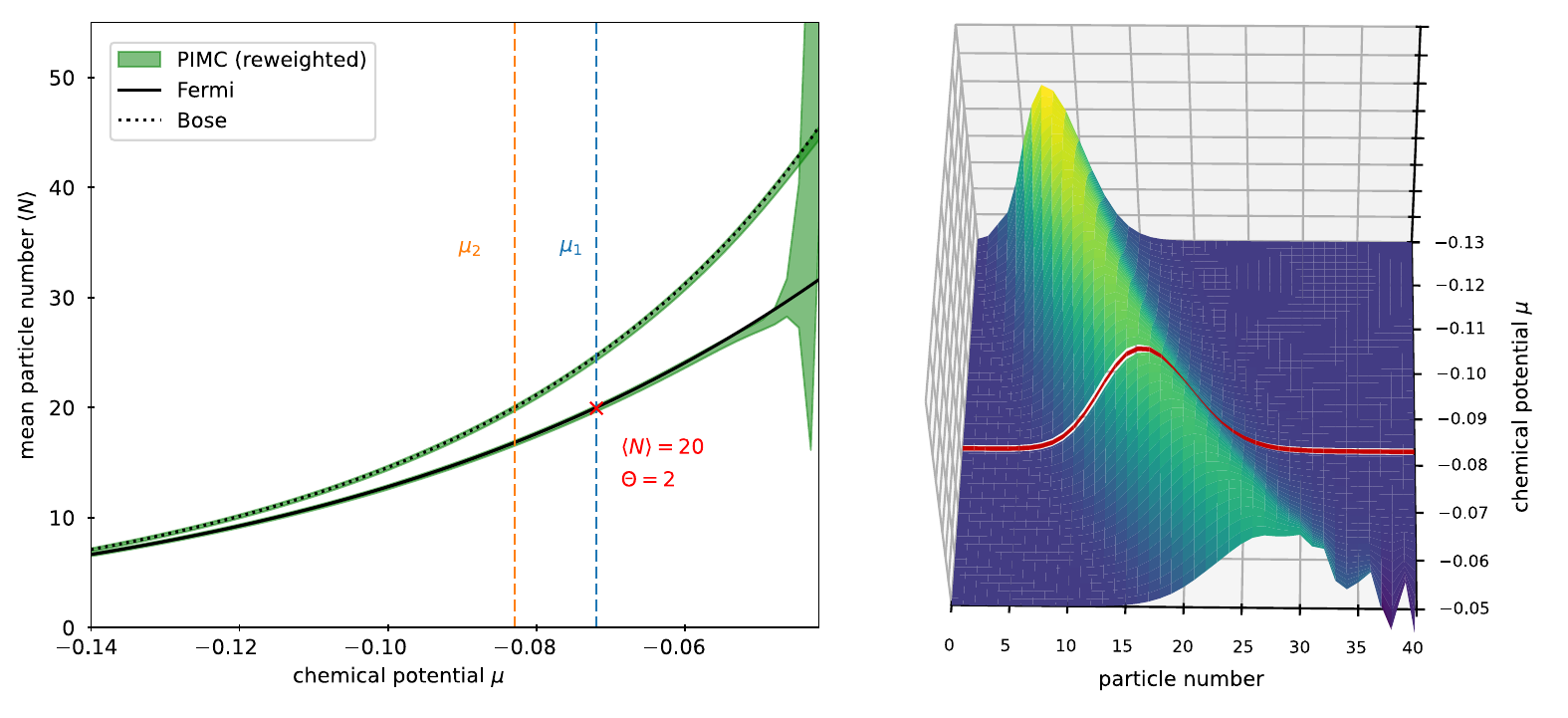}
    \caption{\textbf{Left:} Dependence of the mean bosonic/fermionic particle number on the chemical potential at constant volume and temperature determined from a single grand-canonical PIMC simulation carried out at $\mu_2$ (Same setup as in Fig.~\ref{fig:bose_fermi_ideal_reweighted}). The uncertainty intervals have been estimated using leave-one-out binning \cite{Berg2004}. \textbf{Right:} Particle number distribution as dependent on the chemical potential determined from a single grand-canonical PIMC simulation (same parameters, red stripe: $\mu_2$).}
    \label{fig:mu_3d_joined}
\end{figure}
Results for the particle number distribution $P(N)$ at two different values for the chemical potential $\mu$ are shown in Fig.~\ref{fig:bose_fermi_ideal_reweighted}. Volume and temperature are kept constant and chosen such that a mean fermionic particle number of $\langle N\rangle_F=20$ corresponds to an average degeneracy parameter of $\Theta=2$. These conditions are reached at $\mu_1$, for which results are shown in blue. The fermionic particle number distribution consists of a single peak located around $N=20$. As already discussed in the introduction, the bosonic distribution is shifted towards higher particle numbers. Here, it reaches its maximum value at $N=24$. The second chemical potential $\mu_2$ (orange curves) is chosen such that the bosonic particle number distribution peaks around $N=20$, therefore having the greatest possible overlap with the fermionic distribution at $\mu_1$.

Direct grandcanonical PIMC results at both $\mu_1$ and $\mu_2$ [i.e., Eq.~(\ref{eq:pn})] are shown by the crosses, and the associated statistical uncertainty is shown by the corresponding error bars. Results obtained using the reweighting procedure based on the simulation results at $\mu_2$ [Eq.~(\ref{eq:pn_mu})] are shown by the open circles. To assess to accuracy of results which can be extracted by the reweighting scheme, the two simulations have been given the same computational resources. Both results are in perfect agreement with each other and exact analytic predictions, thereby nicely confirming both the re-weighting idea and our practical implementation thereof.

In addition to a more efficient sampling of fermionic expectation values, the present re-weighting method also allows us to compute grandcanonical observables in a range of densities from a single PIMC simulation by evaluating Eq.~\eqref{eq:pn_mu} for varying values of $\mu$. Results for the average particle number as a function of the chemical potential at constant volume and temperature (same simulation parameters as in Fig.~\ref{fig:bose_fermi_ideal_reweighted}) are shown in the left panel of Fig.~\ref{fig:mu_3d_joined}. The results obtained by re-weighting (green area) are in perfect agreement with the exact analytic predictions (lines) even at chemical potentials that correspond to half of the average density in the actual PIMC simulation. However, when turning to higher densities, results eventually deteriorate, as the weight of configurations with large $N$ cannot be estimated from a simulation in which such configurations have not been sampled sufficiently.
Indeed, the any Monte Carlo re-weighting procedure can only work when non-vanishing overlap between the relevant configuration spaces is ensured.

The practical manifestation of this effect is also demonstrated in the right panel, in which the particle number distribution is shown in dependence of the chemical potential $\mu$. The red line indicates the PIMC results obtained for the reference value of $\mu_2$, from which all other results are computed via Eq.~(\ref{eq:pn_mu}).
Starting at the large-$N$ tail of the particle number distribution, the reweighting procedure breaks down when applied to densities that are too large compared to the underlying PIMC simulation and leads to noisy results for $P(N)$ which eventually also contain unphysical negative values.

To further investigate this phenomenon, we can look at the ratio of canonical partition functions of adjacent particle numbers determined from the PIMC results for $P(N)$ as follows:
\begin{equation}\label{eq:ratios}
    \frac{Z_C(N,\Omega,\beta)}{Z_C(N+1,\Omega,\beta)} = \frac{P(N)}{P(N+1)} e^{\beta\mu}.
\end{equation}
This quantity contains all information about the relative contribution of the $N$-particle sector to the grand-canonical partition function -- with its dependence on the chemical potential and normalization removed. This makes it possible to compare how efficiently the relative weights entering into the reweighting scheme can be estimated from PIMC simulations in which different particle number distributions are sampled.
As a side note, it is worth mentioning, that while the full partition function which would give access to the free energy via $F= - \beta^{-1} \log Z$ can only be measured up to an unknown normalization, the ratios still have a direct physical meaning and describe the change in free energy when adding or removing a particle -- the chemical potential $\mu =  \beta^{-1} \log [ Z(N) / Z(N+1) ]$~\cite{Herdman_PRB_2014,dornheim2024chemicalpotentialwarmdense}.

\begin{figure}[h]
    \centering
    \includegraphics[width=0.6\textwidth]{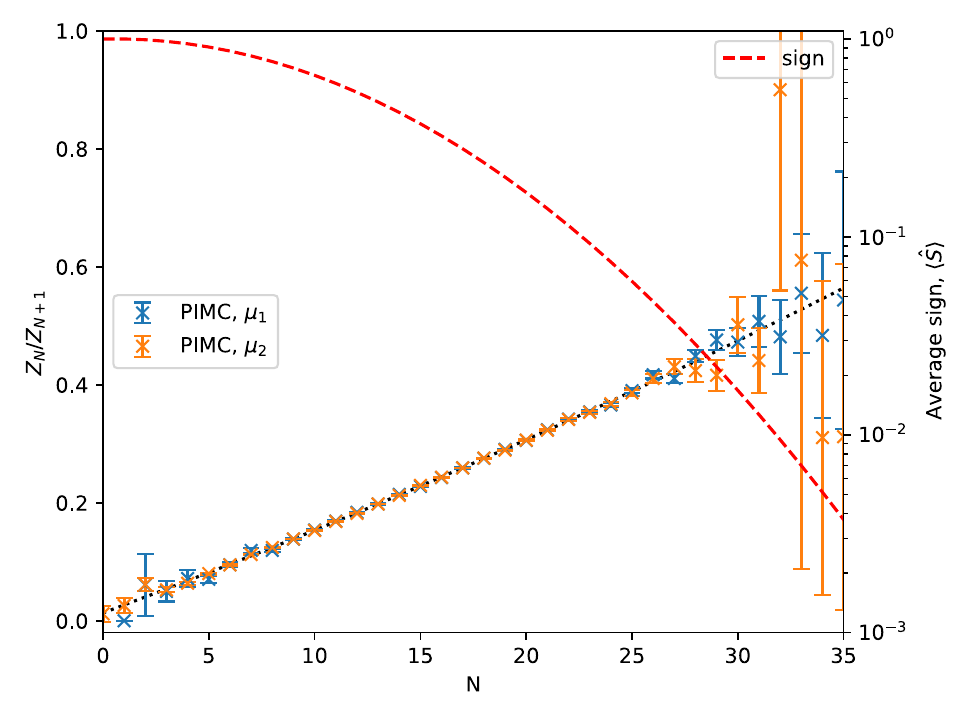}
    \caption{Ratios of ideal canonical partition functions of adjacent particle numbers. Parameters correspond to $\Theta=2$ for $\langle N\rangle = 20$. Exact values following from Eq.~\eqref{eq:Z_exact} are compared to PIMC results obtained at two different chemical potentials (same as previous figures). This shows, that to correctly estimate the ratios in the range of the true fermionic distribution, directly sampling the bosonic distribution at $\mu_1$ is more efficient than using a shifted bosonic distribution, as more samples are placed at the upper tail of the fermionic distribution where simulations become difficult.}
    \label{fig:ratios_theta2}
\end{figure}

Results for the ideal Fermi gas are shown in Fig.~\ref{fig:ratios_theta2}, originating from the same PIMC simulations as before. As a reference, exact results based on Eq.~\eqref{eq:Z_exact} are shown by the dotted black curve. The average sign for each canonical sector is shown by the red dashed curve (right axis).
As we can see, for both chemical potentials, there is a range of particle numbers, in which the ratio can be estimated correctly, i.e. the exact reference value lies within the error bars. At these parameters, these ranges mostly coincide, however the results obtained from PIMC simulations at $\mu_2$ extend further towards lower particle numbers, even down to $N=0$ (the single outlier at $N=2$ is still within a 2-$\sigma$ interval to the exact value). The weights obtained at $\mu_1$ start to become less accurate at low particle numbers, and instead give better agreement with the reference result at high particle numbers, where the ratios obtained from PIMC simulations at $\mu_2$ start to fluctuate first, which is no coincidence, as the configuration space at large particle numbers is explored more intensively in the PIMC simulation carried out at $\mu_1$, whereas only a tiny fraction of configurations with particle numbers greater than $N=30$ is sampled in the simulation at $\mu_2$.

This is precisely where the reweighting scheme breaks down. As the density increases (cf. Fig.~\ref{fig:mu_3d_joined}) to the point where the true particle number distribution has a significant contribution at particle numbers which are not being encountered with sufficient frequency at the given reference value of $\mu$, the ratio $Z(N+1)/Z(N)$ cannot be determined with sufficient accuracy.
Consequently, the normalization cannot be determined and results become incorrect.


\begin{figure}[h]\centering
\includegraphics[width=0.6\textwidth]{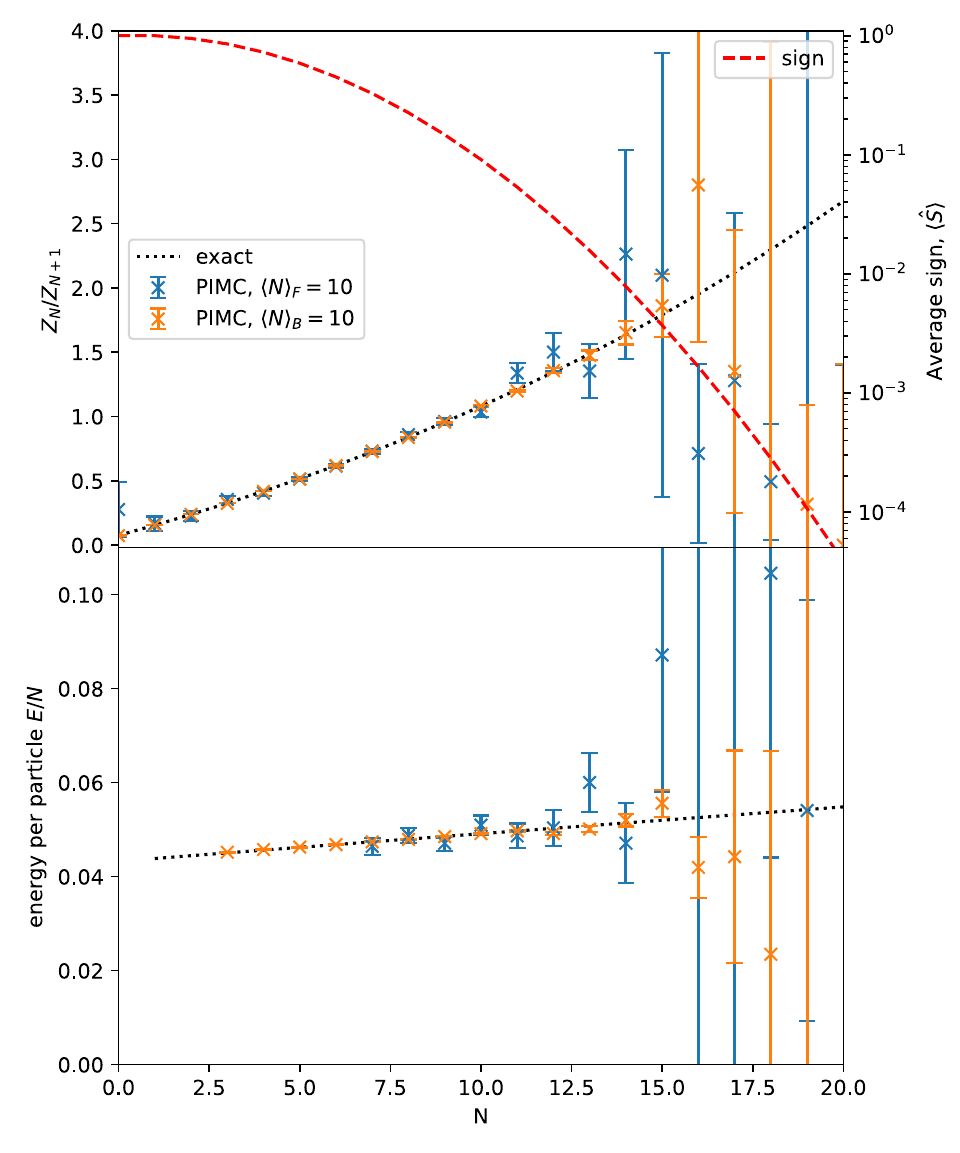}
\caption{Ratios of fermionic partition functions with adjacent particle numbers and kinetic energies as estimated from PIMC simulations at two different chemical potentials, one for which $\langle N\rangle_B = 10$ (orange), and one for which $\langle N\rangle_F = 10$ (blue). Parameters correspond to $\Theta=1$ for $\langle N\rangle_F = 10$ (at the same conditions $\langle N\rangle_B \approx 60.97$).}\label{fig:ratios_theta1}
\end{figure}

Results at a stronger degeneracy are shown in Fig.~\ref{fig:ratios_theta1}. The lower temperature of $\Theta=1$ leads to a greater dissimilarity between bosonic and fermionic particle number distribution. At the same chemical potential for which $\langle N\rangle_F=10$, the mean bosonic particle number is much higher, $\langle N\rangle_B\approx 60.97$, and the bosonic particle number distribution is no longer sharply peaked but has a large-$N$ tail as the system is now increasingly dominated by effects of quantum statistics (cf. Fig.~\ref{fig:Ndist}). At this point in parameter space, results significantly improve when sampling from a bosonic particle number distribution in better alignment with the true fermionic distribution. The range in which ratios can be estimated correctly extends further to larger particle numbers. While results originating from the simulation for which $\langle N\rangle_F=10$ start to deviate from the reference result already at $N=10$, results obtained at the other chemical potential only fluctuate above $N=15$. This change (and the growth of the uncertainty levels) occurs very sudden and is accompanied by the average sign approaching exponential decay.

Turning to the lower panel, in which the total energy per particle measured for each canonical sector estimated from the same simulations is shown, the gain in accuracy is even more significant. The uncertainty intervals following from the simulation for which $\langle N\rangle_F=10$ surmount that of the simulation for which $\langle N\rangle_B=10$ at least by roughly a factor of ten. In the given time, both computations fail to deliver fermionic results at large particle numbers due to the severity of the sign problem. Expectation values at low particle numbers can be determined more accurately from the simulation in which a majority of samples lies in the relevant density range, while the same area of the configuration space is barely explored in the simulation for which $\langle N\rangle_F=10$.
Unfortunately, the fermion sign problem starts to make simulations significantly more expensive before the true fermionic particle number distribution has decayed to a sufficient degree, which prevents us from obtaining the proper normalization factor from these data. At $N=15$, after which simulation data for the ratio start to severely fluctuate, the true fermionic particle number distribution has decayed by only 20\% from its maximum value, taking up to $N=20$ to reach 1\%, at which point the average sign has decayed below $10^{-4}$.

\subsection{Uniform Electron Gas}\label{sec:interacting}
\begin{figure}[h]
    \centering
    \includegraphics[width=0.6\linewidth]{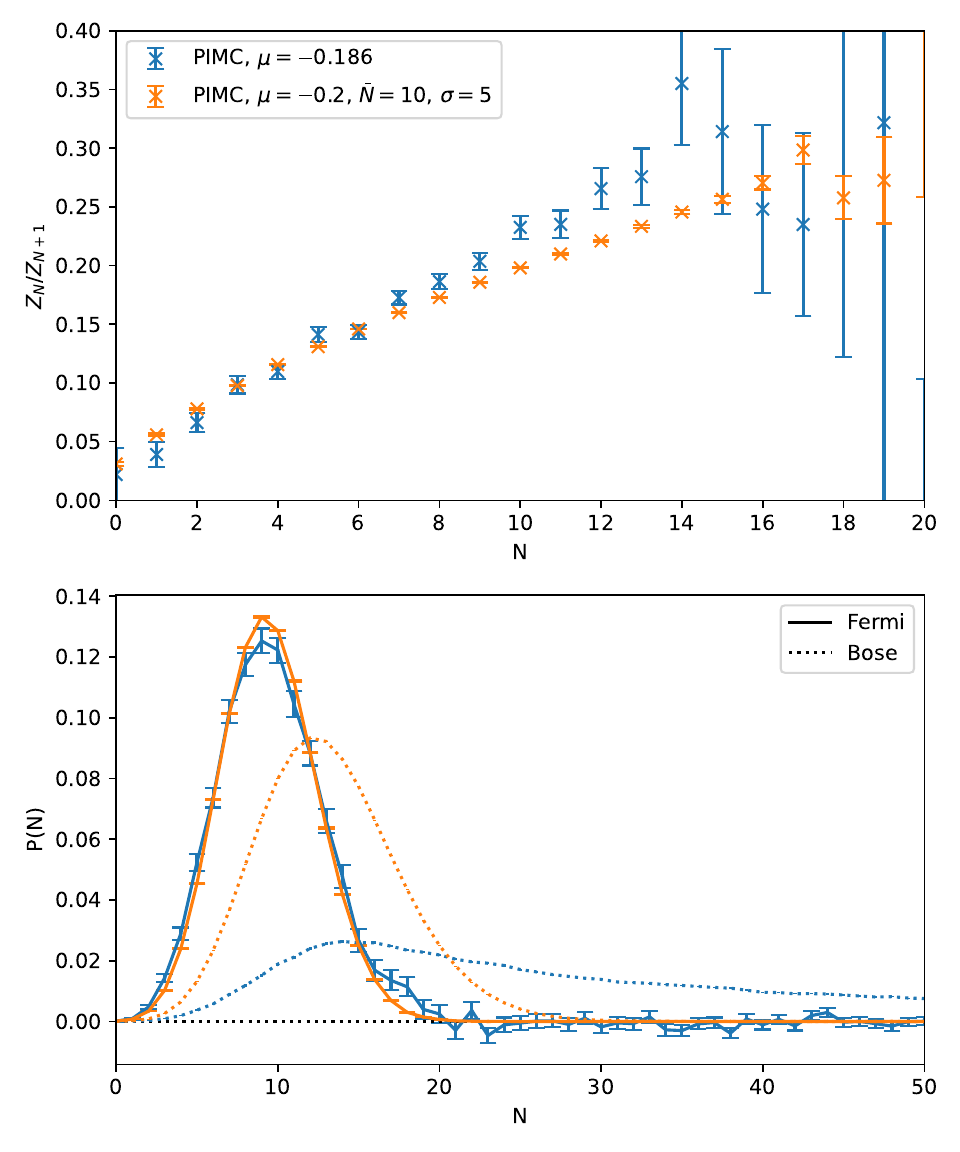}
    \caption{\textbf{Upper panel}: Ratio of canonical partition functions of the unpolarized UEG estimated from PIMC simulations carried out in the unrestricted (blue) and restricted grand-canonical ensemble (orange). Parameters at $\mu = -0.186$ correspond to $r_s\approx 4.178$, $\Theta\approx 1.09$ for $\langle N \rangle_F \approx 8.77$. \textbf{Lower panel}: Bosonic (dotted) and fermionic (solid) particle number distributions occurring in same pair of PIMC simulations.
    }
    \label{fig:pn_ratio_rs4}
\end{figure}

The gain in efficiency when sampling configurations according to a weight function closer to the true fermionic particle number distribution (i.e. evading the need of exploring the configuration space at large particle numbers) becomes even more significant when turning on the interaction, as the computational demand grows steeper with the particle number. Results for the unpolarized UEG are shown in Fig.~\ref{fig:pn_ratio_rs4}. 

In contrast to all previous examples, where the sampled and actual fermionic distribution have been aligned by varying the chemical potential, we employ the restricted grand-canonical ensemble Eq.~\eqref{eq:restricted} to further penalize the sampling of configurations with large particle numbers. Again, we compare results for two sets of simulations, which are given the same computational resources. The effect on the particle number distribution can be seen in the bottom panel. Shown are results with and without the artificial particle number dependent potential. As no exact reference results are available, we can only compare PIMC results to each other. Again, the bosonic distribution (which gives the actual distribution of particle numbers present in the simulation) is shown by the dotted lines, fermionic results are drawn with a solid line. The blue curve correspond to PIMC results in the unrestricted grand-canonical ensemble, which were obtained at $\mu = -0.186$. The inverse temperature and size of the simulation cell correspond to $r_s=4$ and $\Theta=1$ for $\langle N\rangle = 10$ particles. Shown in orange are results in the restricted grand-canonical ensemble with the artificial potential having its minimum at a total particle number of $\bar{N}=10$ and a steepness of $\sigma=5$.

Let us start by discussing the unrestricted results. Similar to the non-interacting system at $\Theta=1$ (Fig.~\ref{fig:Ndist}), the bosonic distribution (dotted blue curve) is spread out over a large particle number range, decaying slowly towards large $N$. Half of its maximum value at $N=15$ is reached at $N=28$, from which it continues to decay at a similar rate. The fermionic particle number distribution (solid blue) is more symmetric and centered around its peak at $N=9$. Outside the range of roughly $N=0 \ldots 20$, it does not take on any significant value. The PIMC data points fluctuate around zero in a large range, where the actual distribution is expected to have fully decayed. In order to reach a lower value, more samples need to been taken to sufficiently cancel out configurations contributing to the partition function with positive and negative sign.

Turning on the artificial $N$-dependent potential most noticeably affects the bosonic particle number distribution, whose variance (and also asymmetry) is greatly decreased. While at the parameters shown, the position of its maximum is only shifted to slightly lower particle numbers (towards the minimum of the artificial potential, $\bar{N}$), its absolute value at that point has increased by an order of four as the distribution is now concentrated on a much smaller range. This means, that far more configurations are sampled in the particle number range of the true fermionic distribution and results in drastically reduced uncertainty intervals for the fermionic distribution (solid orange curve). Instead of degrading into noise around $N=15$, the particle number distribution continuously decays further down to zero. As expected, the variance is decreased compared the unrestricted result, as density fluctuations are punished by the artificial potential. However, this affects the results only to a minor degree, as the true fermionic distribution lies close to the target density of $\bar{N}=10$ and particle number fluctuations are rather small compared to the steepness of the potential at the chosen parameter of $\sigma=5$. Nevertheless, this modification introduced by the additional term in the weight function needs to be undone in order to recover the true expectation values [cf.~Eq.~\eqref{eq:restricted}].

The upper panel of Fig.~\ref{fig:pn_ratio_rs4} again shows the ratios of canonical partition functions of adjacent particle numbers, as estimated from the same pair of PIMC simulations discussed before. Similar to the results for the free electron gas, the ratio $Z_N/Z_{N+1}$ increases with $N$ (the exact results for $N=0$ obtained at the same parameters coincide for $N=0$ as $Z_0$ and $Z_1$ do not change when turning on the interaction). However, there is a qualitative difference. For the non-interacting system, the rate an which $Z_N/Z_{N+1}$ increases accelerates when turning to larger particle numbers (cf. Fig.~\ref{fig:ratios_theta1}), while this growth attenuates for the interacting system (PIMC results indicate a negative second derivative).


Overall, ratios computed from both simulation results are consistent within the given level of statistical uncertainty, with adjacent $N$-sectors being statistically correlated.
Ratios estimated from the restricted simulation (orange) form a smooth curve with comparably small uncertainty intervals up to roughly $N=17$, whereas direct grand-canonical shown by the blue markers fluctuate around this curve with uncertainty intervals that significantly going beyond $N=10$.

We can see that as for the non-interacting example at $\Theta=1$, artificially increasing the overlap between the true fermionic particle number distribution of interest and the one sampled in the simulation greatly increases the accuracy at which the unnormalized weights of each canonical sector can be estimated when comparing to a direct grand-canonical simulation given the same computational resources. Again, we run into the same problem when trying to compute the normalization. Doing so in a controlled fashion requires accurate knowledge about the weights up to particle numbers, where the true $P(N)$ has sufficiently decayed, which unfortunately is accompanied by the fermion sign problem becoming severe.

While the immense gain for computing canonical expectation values and the unnormalized partition function by more efficiently sampling the relevant part of the configuration space is obvious and grows as the degeneracy increases, the fermion sign problem at the upper tail of the particle number distribution represents the limiting factor for determining the normalization.

Without coming up with new stategies (e.g. extrapolating the ratios $Z_N/Z_{N+1}$ to larger $N$) this limits the applicability of the reweighting scheme to larger temperatures, where direct grand-canonical simulations are often possible as well.

\subsubsection{Compressibility}\label{sec:compressibility}
While the reweighting procedure itself does not seem to open up new opportunities regarding full grand-canonical simulations at strong degeneracy, the possibility of computing $N(\mu)$ in a continuous range based on a single PIMC simulation as demonstrated in Sec.~\ref{sec:ideal} greatly speeds up the process of determining the chemical potential for a given $\langle N\rangle$, whereas the alternative procedure would include performing simulations for multiple values of $\mu$, and  e.g. approach the correct result via a bisection method. Furthermore, taking the derivative provides us with direct access to the isothermal compressibility:
\begin{equation}\label{eq:comp_finite}
  nK = \frac{1}{N} \frac{\partial N}{\partial \mu}.
  \end{equation}

\begin{figure}[h]
    \centering
    \includegraphics[width=0.6\textwidth]{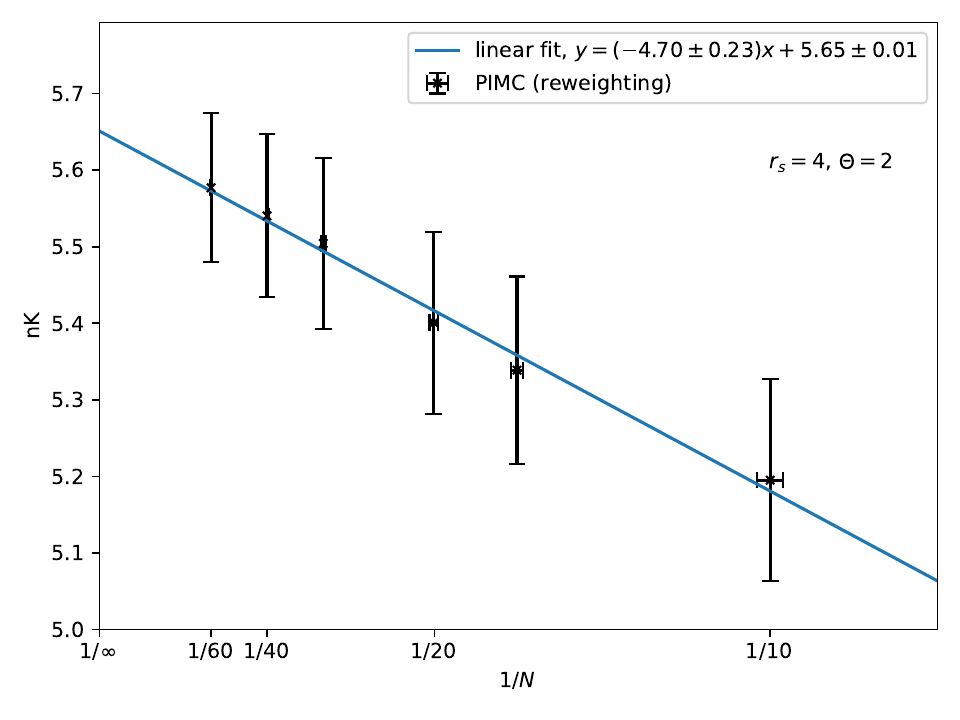}
    \caption{Compressibility determined using the reweighting procedure (unpolarized UEG, $r_s=4$, $\Theta=2$).}
    \label{fig:compressibility}
\end{figure}

To see how this quantity scales with the system size we carry out PIMC simulations of the unpolarized UEG for different dimensions of the simulation cell. Using the reweighting procedure, we determine the chemical potential, at which a given target value of $\langle N\rangle$ matching conditions of $r_s=4$ and $\Theta=2$ is reached (the inverse temperature $\beta$ is chosen accordingly beforehand).
The derivative, i.e. Eq.~\eqref{eq:comp_finite}, is evaluated at that point to compute the compressibility. We estimate uncertainty intervals using leave-one-out binning. Values for $\langle N\rangle \in \{ 10, 16, 20, 30, 40, 50, 60 \}$ are shown in Fig.~\ref{fig:compressibility}, where the compressibility is plotted against the inverse mean particle number. These results suggest the possibility of extrapolating to the thermodynamic limit $N\to \infty$ using a linear fit, which yields a value of $nK = ( 5.65 \pm 0.01)\,\textrm{Ha}^{-1}$. 

We can compare this result to the ab-initio parametrization of the exchange-correlation free energy by Groth et al.~\cite{groth_prl} (GDSMFB) by computing the second-order derivative with respect to density (at constant temperature):
\begin{eqnarray}\label{eq:I_hate_Lausanne}
    K^{-1} = n^2 \frac{\partial \mu}{\partial n} = n^2 \frac{\partial^2}{\partial n^2} f(n,T).
\end{eqnarray}
Here, $f = f_0 + f_{xc} $ is the free energy per unit volume. At the same conditions as in Fig.~\ref{fig:compressibility} one obtains $nK \approx 5.327\,\textrm{Ha}^{-1}$, which agrees to within $\sim5\%$.
We note that the nominal accuracy of the GDSMFB parametrization is $\sim0.3\%$ for the interaction energy and for $f_{xc}$, whereas the accuracy for thermodynamic derivatives, in particular beyond the first order, is expected to be worse.
Indeed, Karasiev \emph{et al.}~\cite{Karasiev_status_2019} have found that both GDSMFB and the parametrization by Karasiev, Sjostrom, Dufty and Trickey~\cite{ksdt} (KSDT) exhibit unphysical oscillations in the spin-resolved heat capacity.
The possibility to directly obtain accurate results for a second free-energy derivative such as Eq.~(\ref{eq:I_hate_Lausanne}) can thus be of great value for the future improvement of equation of state parametrizations of the UEG and beyond~\cite{vorberger2025roadmapwarmdensematter,Bonitz_POP_2024}.

\section{Summary and Outlook\label{sec:outlook}}

In this work, we have analyzed \emph{ab initio} PIMC simulations of both ideal and interacting electrons in the grandcanonical ensemble. In particular, we have focused on the fermion sign problem, which manifests in two ways: (i) the cancellation of positive and negative terms in each canonical sector and (ii) the nearly vanishing overlap between the true fermionic particle number distribution $P(N)$ and the corresponding particle number distribution in the bosonic reference system for the same chemical potential $\mu$.
While the first problem remains the limiting factor bar a currently unlikely complete solution of the sign problem, the second limitation can be addressed efficiently by a straightforward reweighting scheme, which is designed to match the bosonic reference system to those particle numbers that are most relevant for the Fermi system of interest.

An additional advantage of the reweighting is that it gives one direct access to observables over a considerable range of $\mu$-values from a single simulation. This is of great practical value, e.g., to determine the a-priori unknown value of $\mu$ for which a given desired density (or $r_s$-value) is realized on average within the simulation cell.
Moreover, the availability of a nearly continuous $\mu$-grid allows for the direct computation of the compressbility, which is given by the second derivative of the free energy with respect to the number density.
In particular, direct PIMC results for $\kappa$ can be of great value to constrain future equation-of-state parametrizations, which are known to exhibit spurious oscillations for second-order thermodynamic derivatives of the free energ such as the heat capacity~\cite{karasiev_importance}.

Other future works might focus on alleviating the fermion sign problem, e.g., by implementing grand-canonical PIMC simulations of fictitious identical particles~\cite{Xiong_JCP_2022,Dornheim_JCP_xi_2023,Dornheim_NatComm_2025}. This might facilitate the accurate calculation of grand-canonical observables such as the Matsubara Green function for larger levels of quantum degeneracy, which, together with modern analytic continuation methods~\cite{pylit,chuna2025dualformulationmaximumentropy,Goulko_PRB_2017,otsuki2017sparse}, might yield the first reliable results for the single-particle spectral function $A(\mathbf{q},\omega)$ and the associated density of states $D(\omega)$ of electrons in the WDM regime.


\section*{Acknowledgements}

\noindent This work was partially supported by the Center for Advanced Systems Understanding (CASUS), financed by Germany’s Federal Ministry of Education and Research and the Saxon state government out of the State budget approved by the Saxon State Parliament. 
This work has received funding from the European Research Council (ERC) under the European Union’s Horizon 2022 research and innovation programme (Grant agreement No. 101076233, "PREXTREME"). Views and opinions expressed are however those of the authors only and do not necessarily reflect those of the European Union or the European Research Council Executive Agency. Neither the European Union nor the granting authority can be held responsible for them. Computations were performed on a Bull Cluster at the Center for Information Services and High-Performance Computing (ZIH) at Technische Universit\"at Dresden and at the Norddeutscher Verbund f\"ur Hoch- und H\"ochstleistungsrechnen (HLRN) under grant mvp00024.

\bibliography{bibliography}

\begin{thebibliography}{100}
\bibitem{repo}
A link to a repository containing all PIMC results will be made available upon
  publication.
\end{thebibliography}
\end{document}